\documentclass[twocolumn,times,tighten]{aastex63} 

\accepted{September 3, 2019}
\shorttitle{HaloSat}
\shortauthors{Kaaret et al.}

\begin{document}

\title{HaloSat - A CubeSat to Study the Hot Galactic Halo}

\author[0000-0002-3638-0637]{P. Kaaret}
\affiliation{Department of Physics and Astronomy, University of Iowa, Iowa City, IA 52242, USA}
\author{A. Zajczyk}
\affiliation{Department of Physics and Astronomy, University of Iowa, Iowa City, IA 52242, USA}\affiliation{NASA Goddard Space Flight Center, Greenbelt, MD 20771, USA}
\affiliation{Department of Physics, University of Maryland, Baltimore County, 1000 Hilltop Circle, Baltimore, MD 21250}
\author{D. M. LaRocca}
\affiliation{Department of Physics and Astronomy, University of Iowa, Iowa City, IA 52242, USA}
\author{R. Ringuette}
\affiliation{Department of Physics and Astronomy, University of Iowa, Iowa City, IA 52242, USA}

\author{J. Bluem}
\affiliation{Department of Physics and Astronomy, University of Iowa, Iowa City, IA 52242, USA}
\author{W. Fuelberth}
\affiliation{Department of Physics and Astronomy, University of Iowa, Iowa City, IA 52242, USA}
\author{H. Gulick}
\affiliation{Department of Physics and Astronomy, University of Iowa, Iowa City, IA 52242, USA}
\author{K. Jahoda}
\affiliation{NASA Goddard Space Flight Center, Greenbelt, MD 20771, USA}
\author{T. E. Johnson}
\affiliation{NASA Goddard Space Flight Center, Greenbelt, MD 20771, USA}
\author{D. L. Kirchner}
\affiliation{Department of Physics and Astronomy, University of Iowa, Iowa City, IA 52242, USA}
\author{D. Koutroumpa}
\affiliation{LATMOS/IPSL, CNRS, UVSQ Universit\'{e} Paris-Saclay, UPMC Univ. Paris 06, Guyancourt, France}
\author{K.D. Kuntz}
\affiliation{The Henry A. Rowland Department of Physics and Astronomy, Johns Hopkins University, Baltimore, MD 21218, USA}
\author{R. McCurdy}
\affiliation{Department of Physics and Astronomy, University of Iowa, Iowa City, IA 52242, USA}
\author{D. M. Miles} 
\affiliation{Department of Physics and Astronomy, University of Iowa, Iowa City, IA 52242, USA}
\author{W. T. Robison}
\affiliation{Department of Physics and Astronomy, University of Iowa, Iowa City, IA 52242, USA}
\author{E. M. Silich}
\affiliation{Department of Physics and Astronomy, University of Iowa, Iowa City, IA 52242, USA}


\correspondingauthor{Philip Kaaret}
\email{philip-kaaret@uiowa.edu}

\begin{abstract}

HaloSat is a small satellite (CubeSat) designed to map soft X-ray oxygen line emission across the sky in order to constrain the mass and spatial distribution of hot gas in the Milky Way. The goal of HaloSat is to help determine if hot gas gravitationally bound to individual galaxies makes a significant contribution to the cosmological baryon budget. HaloSat was deployed from the International Space Station in July 2018 and began routine science operations in October 2018. We describe the goals and design of the mission, the on-orbit performance of the science instrument, and initial observations.

\end{abstract}


\section{Introduction} \label{sec:intro}

Astronomy at wavelengths that do not penetrate Earth's atmosphere requires space-borne observatories which tend to be costly, severely limiting their number. As of early 2018, there were only seven X-ray observatories in orbit as compared to the $\sim$100 ground-based optical telescopes with apertures of 2~m or larger. The development of the CubeSat standard \citep{Hevner2011} decoupled the design of small satellites from specific launch opportunities and has led to a large increase in the frequency of small satellite flights including frequent launch opportunities for scientific missions \citep{Crusan2019}. The commercialization of small satellite technologies has both increased the capabilities of small satellites and decreased their cost. Small spacecraft with the pointing accuracy, communications, and power resources needed for modest astronomical instruments can now be purchased at prices similar to that of a 1-m optical telescope. This enables low-cost, space-borne astronomical observatories.

HaloSat is the first astrophysics-focused CubeSat mission funded by NASA's Astrophysics Division. HaloSat's scientific goal is to constrain the mass and spatial distribution of hot gas associated with the Milky Way by mapping soft X-ray line emission from highly ionized oxygen in order to determine if hot halos associated with individual galaxies make a significant contribution to the cosmological baryon budget.  The mission was developed on a rapid time scale, less than 2.5 years from the start of funding to launch, and at modest cost, less than \$4M.  HaloSat was deployed from the International Space Station (ISS) on 13 July 2018 and began routine science operations in October 2018.

In the following, we give an overview of the HaloSat mission including the science goals and mission design in section~\ref{sec:overview}. We describe the science instrument in section~\ref{sec:instrument} and the spacecraft and operations in section~\ref{sec:spacecraft}. We present the on-orbit performance in section~\ref{sec:performance}. We discuss initial halo observations and plans for the survey in section~\ref{sec:halo}. Some of the text and figures in this paper have appeared previously in conference proceedings \citep{Zajczyk2018, Kaaret2019}.


\section{Mission Overview} \label{sec:overview}

\subsection{Science Goals}

Our Milky Way galaxy provides a local laboratory for understanding the missing baryon problem. The total mass of the Milky Way has been dynamically measured to be $(1.0-2.4) \times 10^{12} M_{\odot}$ \citep{Boylan2013}. If the cosmological baryon fraction of 15.5\% \citep{Planck2014} applies on the scales of individual galaxies, then the Milky Way's baryonic mass should be $(1.6-3.7) \times 10^{11} M_{\odot}$. However, the observed baryonic mass in stars and disk gas is only $0.7 \times 10^{11} M_{\odot}$ \citep{McMillan2011,Dame1993}. Thus, more than half the expected baryons in the Milky Way are missing. 

The presence of a hot Galactic halo has long been suspected \citep{Spitzer1956}. Absorption lines at zero redshift in the X-ray spectra of extragalactic objects firmly establish the existence of an extended distribution of hot gas, $\sim$10$^{6}$~K, associated with the Milky Way \citep{Nicastro2002}. Other evidence for a hot halo includes ubiquitous gas depletion in dwarf galaxies within 270~kpc of the Milky Way \citep{Grcevich2009} understood as ram-pressure stripping and the dispersion measures of pulsars in globular clusters and the Magellanic clouds \citep{Zhezher2017}.

X-ray absorption line measurements of the halo are feasible along only about 30 lines of sight. In contrast, X-ray emission lines can be measured in any direction and provide a means to obtain a synoptic view of the halo gas. Data from existing X-ray observatories, primarily {\it XMM-Newton} and {\it Suzaku}, have been used for emission line measurements of the halo \citep{Smith2007,Yoshino2009,Henley2010,Henley2012,Miller2015}. However, the existing X-ray observatories are not designed for the efficient study of large-angular-scale diffuse emission and the spectra are often contaminated by heliospheric foreground emission that limits the accuracy of measurements of the halo \citep{Slavin2013}.

HaloSat's scientific goal is to constrain the mass and spatial distribution of hot gas associated with the Milky Way. We chose to do this by measuring line emission from oxygen, the most cosmologically abundant element that is not fully ionized at the temperatures present in the halo. Oxygen in the halo is highly ionized and the strongest emission is in the soft X-ray range from the \ion{O}{7} triplet near 574~eV and the pair of \ion{O}{8} lines near 654~eV. To advance over previous studies of halo oxygen line emission, we set an observational goal of achieving a 1$\sigma$ statistical accuracy of $\pm 0.5$~LU (LU $= \rm photons \, cm^{-2} \, s^{-1} \, ster^{-1}$) on the sum of the \ion{O}{7} and \ion{O}{8} line emission in the 500-700 eV range for an oxygen line strength of 5 LU.

The Milky Way's halo fills the entire sky. The observational goal of HaloSat is to survey at least 75\% of the sky with priority given to fields at Galactic latitudes $|b| \ge 30\arcdeg$. Modest angular resolution of 15$\arcdeg$ or less is sufficient to map the halo emission. HaloSat uses non-imaging detectors to minimize cost and complexity, hence the angular resolution is set by the field of view.  We chose a large field of view with a $10\arcdeg$ diameter full response dropping to zero response at a diameter of $14\arcdeg$, which is consistent with the required angular resolution.

\subsection{Mission Design}

We chose to implement HaloSat as a CubeSat.  The largest CubeSats being regularly flown by NASA at the time of mission formulation were in the ‘6U’ form factor with a volume of 8600~cm$^{3}$. Approximately one third of the volume was allocated to the spacecraft structure and systems leaving a volume of roughly 6000~cm$^{3}$ available for the science instrument. The choice of a CubeSat also limits the mission duration. The most common launch for US-based CubeSats is to the ISS and the relatively low altitude, $\sim$400~km limits the orbital lifetime. HaloSat's orbital lifetime is expected to be 2.5 years. Furthermore, CubeSat electronics often use commercial rather than radiation hardened electronic components to minimize cost which can lead to premature failure in the space environment. HaloSat was designed with a baseline mission lifetime of 405 days based on the longest lifetimes achieved with similar components in previous missions.

HaloSat is designed to be sensitive to diffuse X-ray emission. The number of oxygen line photons detected from a diffuse source is $N = f A \Omega T$, where $f$ is the diffuse line flux generally measured in LU, $A$ is the detector effective area at the oxygen line, $\Omega$ is the detector/telescope field of view, and $T$ is the observation time. The figure of merit for observing diffuse emission is $A \Omega$ or `grasp'. Due to the CubeSat volume constraints, HaloSat uses three small detectors that view the sky through mechanical collimators with no optics. Each detector has an effective area for X-rays of 600~eV of about 5.1~mm$^{2}$. However, HaloSat's field of view is near 100 square degrees, enabling it to efficiently survey the sky. The total grasp of HaloSat at 600~eV is $\rm 17.6 \, cm^{2} \, deg^{2}$. The grasp of the Chandra X-ray Observatory at the same energy was about $2 \times$ lower at launch, while the total grasp of the two {\it XMM-Newton} EPIC-MOS is about $4 \times$ larger. Thus, a CubeSat can be competitive with major space observatories when designed for a specific goal, such as survey efficiency.

The accuracy of current emission line measurements of the halo is limited by foreground oxygen emission produced by solar-wind charge exchange (SWCX), when energetic particles in the solar wind exchange charge with neutral atoms within the solar system \citep[for a review, see][]{Kuntz2019}. HaloSat observes towards the anti-Sun direction during the nighttime half of the 93-minute orbit of the spacecraft around Earth to minimize this foreground. This is not possible with {\it XMM-Newton} because it has a fixed solar array that restricts observations to a Sun angle range of 70$\arcdeg$-110$\arcdeg$. Because the solar system background presents the dominant uncertainty in current line intensity measurements, HaloSat has an additional science goal to improve our understanding of SWCX emission and conducts observations specifically devoted to this goal.

\subsection{Solar wind charge exchange emission}

Solar wind charge exchange (SWCX) emission occurs when a highly charged ion of the solar wind interacts with a neutral atom and gains an electron. The resulting ion is in an excited state and may decay by emitting an X-ray. SWCX emission is produced within Earth's magnetosheath and throughout the heliosphere. The line flux is the integral over the line of sight of the product of the ion density, the neutral density, the relative velocity between the ion and neutral, the charge exchange cross-section, and the line emission probability. The magnetosheath responds rapidly to changes in the solar wind flux, so its SWCX emission is strongly time-variable. The heliospheric emission is integrated over a long line of sight, sampling about a month of solar wind conditions, and varies more slowly.


The distribution of heliospheric emission is determined by the geometry of the target gas. Neutral interstellar gas flows at $\sim 25$~km/s through the Solar System. This gas, mostly hydrogen but with $\sim 15$\% helium, flows from the Galactic direction $(l \sim 3\arcdeg, b \sim 16\arcdeg)$, placing the Earth downstream of the Sun in early December. The flow of interstellar hydrogen is affected by both radiation pressure and gravity, and the hydrogen becomes strongly ionized through charge exchange with solar protons and photo-ionization so that the hydrogen is denser upstream of the Sun than downstream. In contrast, the interstellar helium flow is not strongly ionized but is affected mainly by gravity, which focuses the flow downstream of the Sun into the `He-focusing cone'.

The heliospheric \ion{O}{7} and \ion{O}{8} emission are calculated from the interstellar neutral H and He distributions and measurements of the solar wind provided by solar and heliospheric observatories \citep{Koutroumpa2009}. A crucial input to this modeling is knowledge of the O-He interaction cross section \citep{Galeazzi2014}. HaloSat observed along the He-focusing when the Earth passed through the He-focusing cone in December 2018 with observations made at roughly monthly intervals from two months before the passage to two months after. By correlating the observed soft X-ray emission with the He distribution along the line of sight, these HaloSat observations will provide an accurate measurement of the O-He cross section and an absolute scale for the SWCX models. These observations cannot be done by any other current observatory.  

HaloSat is also carrying out a series of observations to study magnetospheric SWCX emission. A selected set of targets is observed multiple times, once near the anti-Sun direction in order to minimize magnetospheric SWCX emission and also as the target enters and exits the anti-Sun hemisphere and is viewed through the flanks of the magnetosheath. These flank observations should have significant magnetospheric SWCX emission. By measuring the difference between the anti-Sun versus the flank observations, we will be able to characterize the magnetospheric SWCX emission and its dependence on the solar wind and viewing angle through the magnetosphere.

Optimizing the epoch of each halo observation should significantly reduce the SWCX contribution for HaloSat observations of the halo. The dedicated heliospheric and magnetospheric SWCX observations should improve the accuracy with which we can model the remaining SWCX emission and, thus, improve the accuracy of HaloSat measurements of the Milky Way's halo.


\begin{figure}[tb]
\includegraphics[width=3.25in]{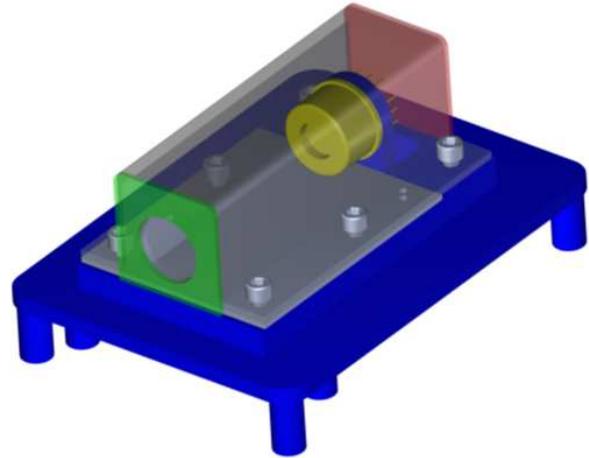}
\caption{Schematic of an X-ray detector assembly \citep{Zajczyk2018}. The SDD is shown in gold.  The copper-tungsten passive shield is shown in gray with the top part semi-transparent, the forward end cap in green with the sky aperture, and the rear end cap in red.  The detector is mounted on an aluminum baseplate shown in blue with the front-end electronics (not shown) mounted on the bottom of the plate.
\label{fig:detector}}
\end{figure}

\section{Science Instrument} \label{sec:instrument}

The science instrument of HaloSat consists of three identical detector units each containing an X-ray detector assembly and signal processing electronics \citep{Zajczyk2018}. The X-ray detectors are silicon drift detectors (SDDs) from Amptek, Inc. Each SDD has an active area of 17~mm$^2$ and is in a sealed package along with a multilayer collimator and a thermoelectric cooler package viewing the sky through a $\rm Si_3 N_4$ window covered with a thin layer of aluminum. The SDD is cooled to -30 C during operation, but the window remains at ambient temperature which should help reduce the buildup of contaminants. The SDD provides no imaging capability.

\begin{figure}[tb]
\includegraphics[width=3.25in]{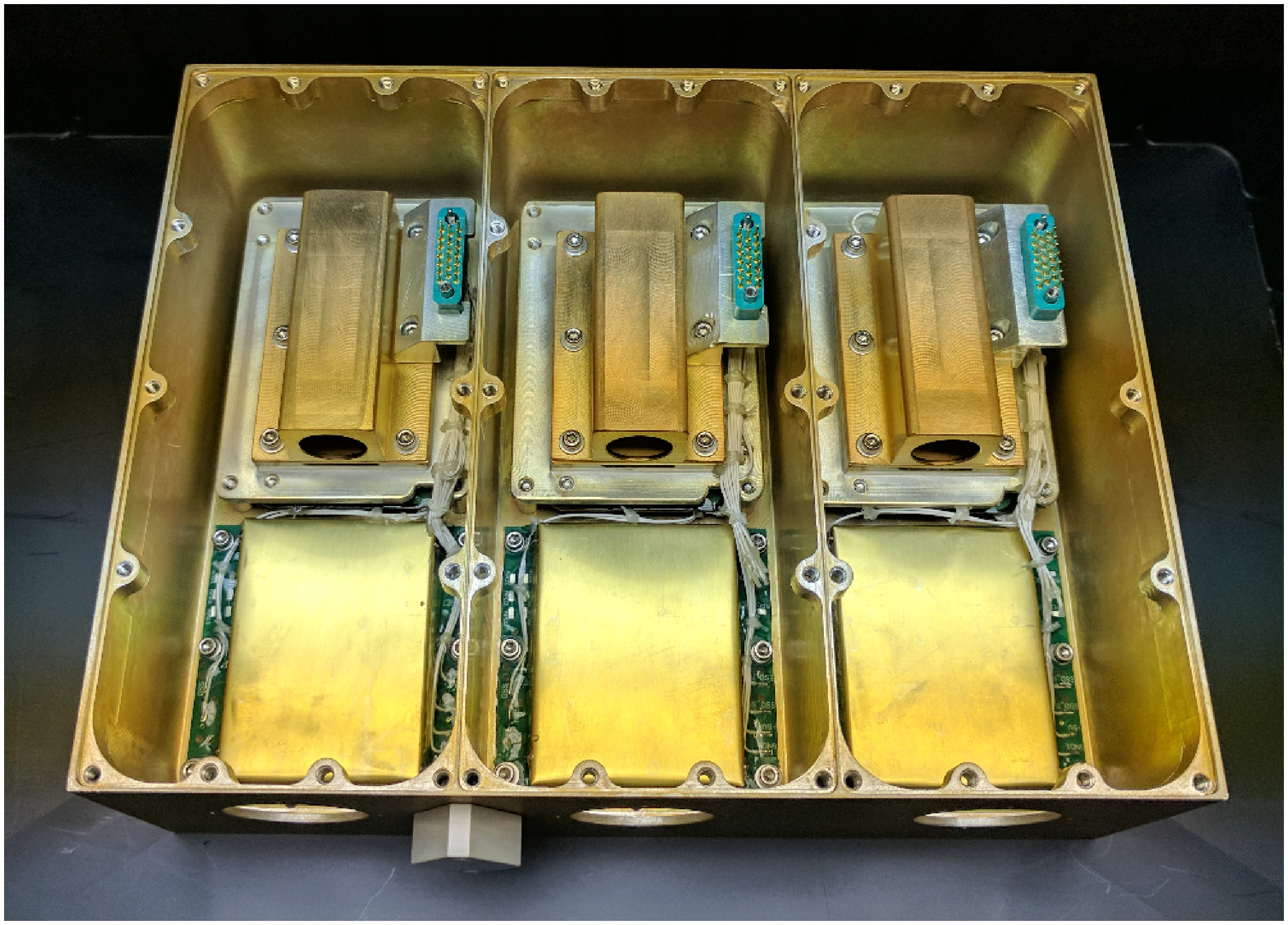}
\includegraphics[width=3.25in]{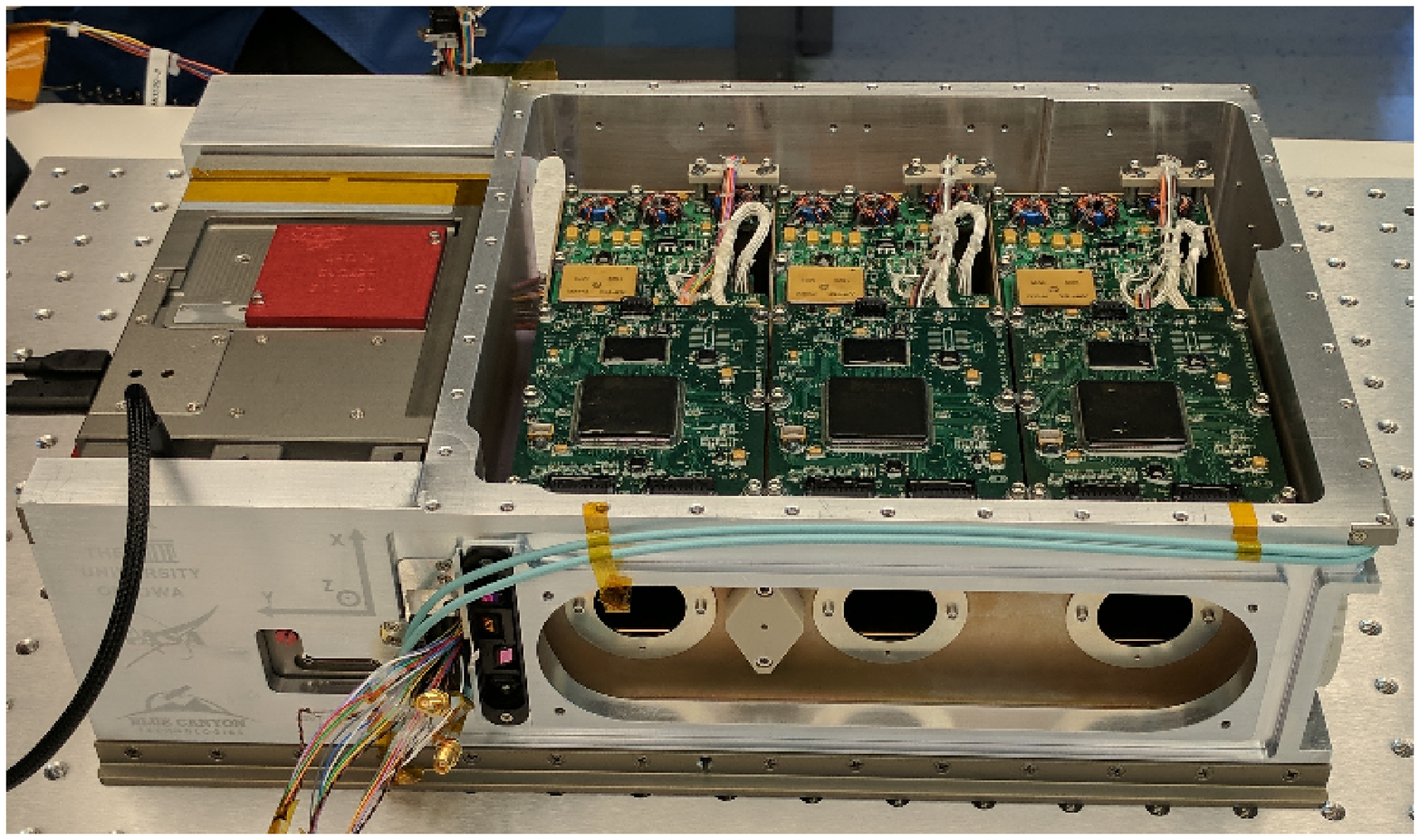}
\caption{Top - The three flight detectors mounted in the instrument chassis. The passive shields are visible towards the back; the detectors are inside. The brass housings towards the front are high-voltage power supplies. The X-ray apertures are at the front. Bottom - Science instrument mounted in the flight bus chassis. Printed circuit boards for analog electronics and the DPUs are on the top of instrument chassis with the alignment washers and a cover over a flat mirror used for optical alignment at the front.  The instrument cover and solar array are not attached. \label{fig:instrument_bus}}
\end{figure}

Each detector is mounted in a compartment inside the instrument chassis made of aluminum. To minimize background from charged particle interactions and the diffuse X-ray background, each SDD is surrounded by a shield made of 1.2~mm thick copper-tungsten metal matrix composite electroplated with a 2.5~$\mu$m layer of nickel and an outer 1.3~$\mu$m layer of gold, see Fig.~\ref{fig:detector}. The shield has a circular aperture through which the SDD views the sky, see Fig.~\ref{fig:instrument_bus}. There is a 0.78~mm thick aluminum washer at the front of each detector compartment with a circular aperture that defines the field of view (FoV). The full-response radius was measured in ground testing to be $5.02\arcdeg$ and the zero-response radius to be $7.03\arcdeg$ with a linear decrease between. The response-weighted effective field of view is 0.0350~steradians.

Each X-ray creates a charge pulse. Pulses triggering a lower level discriminator are digitized and the pulse height and time of arrival, with a resolution of 0.05~s, are recorded by a data-processing unit (DPU) which is a field-programmable gate array programmed with a microprocessor core. We refer to the detector units using numbers encoded into their DPUs which are 14, 54, and 38 as viewed from left to right in Fig.~\ref{fig:instrument_bus}.

The X-ray energy to pulse height conversion was measured during ground calibration with the detectors illuminated by an X-ray beam with fluorescence emission at the F~K$\alpha$ line (676.8 eV) from a Teflon target along with lines from Al, Si, Cr, and Fe, and by a $^{55}$Fe radioactive source \citep{Zajczyk2018}. Measurements were made at instrument temperatures ranging from $-25$~C to $+40$~C. The pulse height to X-ray energy conversion is linear with a non-zero offset. The offset is a linear function of temperature while the slope is a quadratic function of temperature.  The energy scale shifts are up to $\sim$15~eV across the range from $-25$~C to $+40$~C. Details of the ground calibration are presented in \citet{Zajczyk2019}. The energy resolution averaging over all temperatures was measured to be $88.3 \pm 3.5$, $84.3 \pm 2.8$, and $82.0 \pm 1.4$ eV (FHWM) at the F~K$\alpha$ line and $138.5 \pm 2.1$, $136.5 \pm 0.8$, and $137.3 \pm 1.9$ eV (FWHM) at the Mn~K$\alpha$ line (the intensity-weighted average of the K$\alpha_1$ and K$\alpha_2$ lines is 5895.0~eV) for DPUs 14, 54, and 38, respectively.


\begin{figure}[tb]
\includegraphics[width=3.25in]{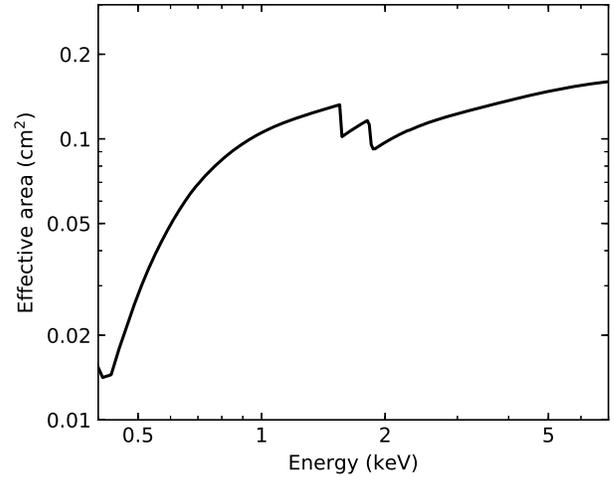}
\caption{Effective area versus energy of a single HaloSat SDD.
\label{fig:arf}}
\end{figure}

Detector response matrices were prepared using software that models the response of silicon detectors \citep{Scholze2009}. The software was modified for the SDDs used for NASA's Neutron star Interior Composition Explorer (NICER) instrument \citep{Gendreau2016} and kindly provided to us by Dr. Jack Steiner of MIT. The NICER SDDs are identical to those on HaloSat except for use of a thinner window. We adjusted the relevant detector parameters using the ground calibration data and information on the windows supplied by Amptek, Inc., and HS-Foils, Oy. The effective area of a single HaloSat SDD is shown in Fig.~\ref{fig:arf}. The response files are compatible with the Xspec spectral fitting software \citep{Arnaud1996} which is commonly used in X-ray astronomy to enable use of HaloSat data by the astronomical community.

\section{Spacecraft and Operations} \label{sec:spacecraft}

To minimize development costs and enhance the probability of mission success by the use of flight-proven components, we chose to use a commercial CubeSat `bus' provided by Blue Canyon Technologies, Inc., (BCT) to provide attitude control, command and data handling, and a power system. Power is provided by a deployable solar array that charges the on-board batteries during the day side of the spacecraft orbit. The spacecraft can be slewed at a 2$\arcdeg$ per second rate and has a pointing accuracy of $\pm 0.002\arcdeg$ (1$\sigma$) \citep{Hegel2016}. An on-board CADET radio is used to down-link telemetry to and receive commands from a radio ground station at NASA's Wallops Flight Facility. A GlobalStar radio provides occasional housekeeping information.

Blue Canyon Technologies performs mission operations based on observation plans generated at the University of Iowa (UI). All the instrument telemetry (which includes X-ray event data, instrument housekeeping data, and spacecraft attitude and orbit information) is captured in a database and processed into a set of FITS (Flexible Image Transport System) format files including spectra and event lists for each target. The FITS files will be archived at the High Energy Astrophysics Science Archive Research Center (HEASARC) and made publicly available within 5 months after mission completion.

\clearpage 
\section{On-Orbit Performance} \label{sec:performance}

The first observations with HaloSat were done to measure the instrument pointing and field of view and to verify the spectral response and effective area.  

\begin{figure}[tb]
\includegraphics[width=3.25in]{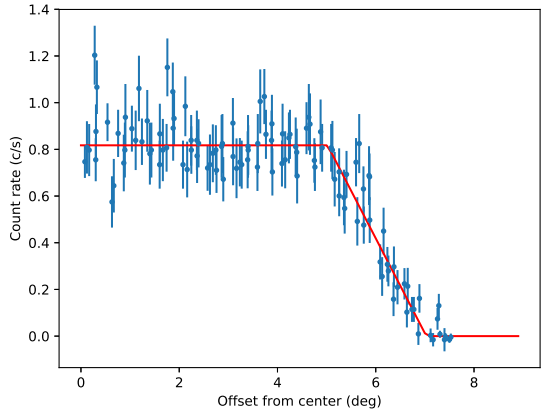}
\caption{X-ray count rate versus pointing offset from the Crab for DPU 54.
\label{fig:crab_offset}}
\end{figure}

\begin{table}[tb]
\caption{Pointing offsets.\label{tab:crab_offset}}
\begin{center}
\begin{tabular}{lcc} \tableline
DPU &    $X$ (deg) &  $Y$ (deg) \\ \tableline
14  &    -0.11   &  0.18 \\
54  &    -0.01   &  0.15 \\
38  &    -0.05   &  0.10 \\ \tableline
\end{tabular}
\end{center}
\end{table}

\subsection{Pointing and Field of View}

The Crab is a pulsar wind nebula powered by a young pulsar with a spin period of about 33 ms.  The Crab has been used as a calibration target since the early days of X-ray astronomy \citep{Toor1974}.  We used the Crab to measure the alignment between the boresights of the X-ray instruments and the coordinate system defined by the star trackers on the spacecraft bus.

We performed a series of slew maneuvers in which the science instrument was pointed towards the Crab and then the pointing was gradually offset while the spacecraft roll angle was held fixed.  Eight different maneuvers were performed corresponding to eight different roll angles at equal intervals in the spacecraft frame. The X-ray count rate versus offset data were fit to a model matching the FoV measured on the ground with the center being a fit parameter. Good fits were obtained, see Fig.~\ref{fig:crab_offset}, verifying the ground FoV measurements. We found an offset of about $1.0\arcdeg$ in the spacecraft Y direction from the nominal pre-flight instrument boresight.  This correction was applied to the pointing of observations obtained after 1 December 2018.  

After the correction, another pointing test was performed and the best fit FoV center was found to be consistent with the expected position within $\pm 0.11\arcdeg$ in the spacecraft $X$ direction and $\pm 0.18\arcdeg$ in the spacecraft $Y$ direction for all DPUs, see Table~\ref{tab:crab_offset}. The count rate for DPU 54 versus radial offset from the best fit center for the best fit model is shown in Fig.~\ref{fig:crab_offset} and the pointing offsets for each detector are presented in Table~\ref{tab:crab_offset}. The uncertainties on the offsets are $\pm 0.05\arcdeg$ (1$\sigma$). We conclude that the pointing of the X-ray boresight of HaloSat is accurate to within $\pm 0.23 \arcdeg$, which is a small fraction of the FoV. The X-ray pointing uncertainty is dominated by the accuracy to which we are able to measure the relative alignment between the X-ray detectors and the spacecraft reference frame. The median offset between the commanded target position during observations and the spacecraft pointing measured by the attitude control system is 0.0007$\arcdeg$.

\subsection{Spectral Response}

To check the on-orbit X-ray energy scale calibration, we examined spectra obtained while observing the dark side of the Earth and while observing the supernova remnant (SNR) Cassiopeia A. 

\begin{figure}[tb]
\includegraphics[width=3.25in]{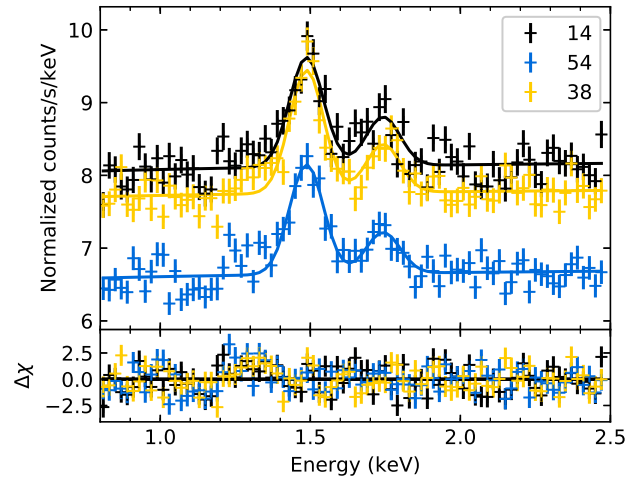}
\caption{X-ray spectra of the dark Earth. Data from all three detectors are shown as indicated by the DPU numbers in the legend. Emission lines are visible from neutral Al at 1.49~keV and neutral Si at 1.74~keV.
\label{fig:dark_earth_spec}}
\end{figure}

The dark Earth observations show instrumental lines from aluminum and silicon likely due to fluorescence by energetic particles. Data were processed using the temperature-dependent ground energy calibration and filtered to maximize the significance of these lines.  The resulting spectra were fit with a model consisting of two Gaussians and a powerlaw, see Fig.~\ref{fig:dark_earth_spec}. With the line centroid energies fixed to the average of the laboratory values of the $K\alpha_1$, $K\alpha_2$, and $K\beta$ lines weighted by their relative strengths (1.4875~keV for Al and 1.7425~keV for Si), we obtained a good fit with $\chi^2$/DoF = 337.7/239. Allowing the line centroid energies to vary did not produce a significant improvement in the fit with an F-test probability of 0.35. The line centroid error ranges include the weighted laboratory values and the best fit centroids for the Al line are all within 3~eV of the laboratory value.

\begin{figure}[tb]
\includegraphics[width=3.25in]{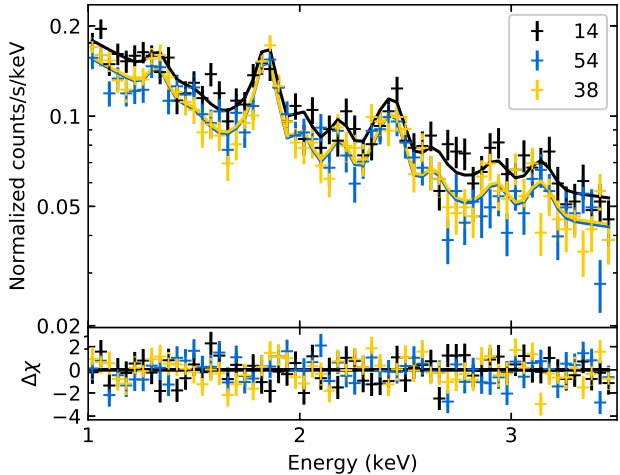}
\caption{X-ray spectra of the Cas A field in the 1-3.5~keV band. Data from all three detectors are shown as indicated by the legend. Prominent emission lines are visible from Si XIII at 1.86~keV and S XV at 2.45~keV. Table~\ref{tab:casa_lines} shows the lines used in fitting the spectra.
\label{fig:CasA}}
\end{figure}

\begin{table}[tb]
\caption{X-ray lines from Cas A.\label{tab:casa_lines}} \vspace{-8pt}
\begin{center}
\begin{tabular}{lcc} \tableline
Line & Energy (keV) & EW (keV) \\ \tableline
Mg He$\alpha$ & 1.3375 & 0.04 \\ 
Si He$\alpha$ & 1.8558 & 0.40 \\ 
Si Ly$\alpha$ & 2.0053 & 0.13 \\ 
Si He$\beta$  & 2.1830 & 0.13 \\ 
Si Ly$\beta$  & 2.3770 & 0.06 \\ 
S He$\alpha$  & 2.4510 & 0.17 \\ 
S Ly$\alpha$  & 2.6220 & 0.09 \\ 
S He$\beta$   & 2.9218 & 0.09 \\ 
Ar He$\alpha$ & 3.1400 & 0.13 \\ 
\tableline
\end{tabular}
\end{center}
\end{table}

Cas A has strong emission lines from heavy elements in its X-ray spectrum. X-ray emission lines from Mg, Si, S, and Ar were first detected with the solid-state spectrometer on Einstein \citep{Becker1979} and first mapped with ASCA \citep{Holt1994}. Cas A has been used to calibrate the energy scale of previous X-ray instruments \citep{Jahoda2006}.

The HaloSat field centered on Cas A includes another SNR, CTB 109, and several point sources, but the emission is dominated by Cas A.  We extracted spectra of the Cas A for all three DPUs, see Fig.~\ref{fig:CasA}, and fit them in the 1.0-3.5 keV range with a model consisting of a powerlaw and nine Gaussians for the astrophysical emission and a powerlaw with photon index fixed to 0.65 for the instrumental background that was not modified by the response matrix.  The parameters of the astrophysical model were set equal for the different detectors while the normalization of the instrumental background was allowed to vary between detectors. Line energies were extracted from the AtomDB database of atomic transitions and centroids for blends were calculated from their relative intensities and are given in Table~\ref{tab:casa_lines}. 

The continuum X-ray spectrum of Cas A is typically described as the sum of two thermal plasma components and a powerlaw, but a single powerlaw produces an adequate fit over the limited energy band used in the fit with $\chi^2$/DoF = 438.31/358. The best fit equivalent widths (EW) for the detected lines are presented in Table~\ref{tab:casa_lines}. The line widths are consistent with the energy resolution measured during the ground calibration. Allowing the slope or the intercept of the channel to energy conversion to vary did not significantly improve the fit with F-test probabilities of 0.20 and 0.16, respectively.  The best fit response slopes differ by less than 0.2\% from the ground calibration while the best fit intercepts differ by less than 5~eV.

\subsection{Effective Area}

\begin{figure}[tb]
\includegraphics[width=3.25in]{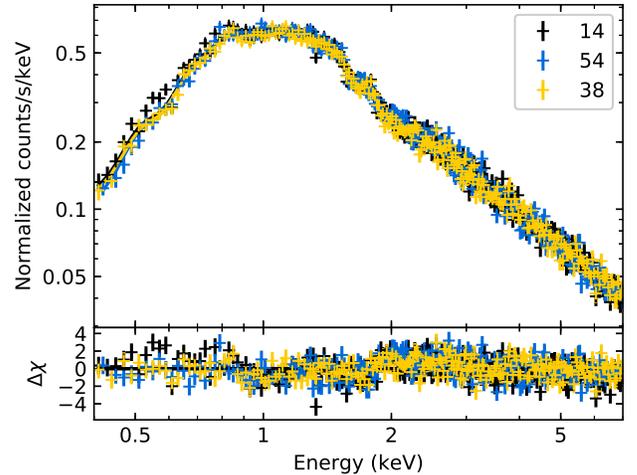}
\caption{X-ray spectra of the Crab in the 0.5-7~keV band. Data from all three detectors are shown as indicated by the legend.
\label{fig:crab_spec}}
\end{figure}

\begin{table}[tb]
\begin{center}
\caption{Crab flux from various instruments.\label{tab:crab_flux}}
\begin{tabular}{lclllc} \tableline
Instrument  & Norm           & $\Gamma$        & $N_H$         & Flux  \\ \tableline
Historical  &   9.7$\pm$1.0  & 2.100$\pm$0.030 & 3.54          & 11.0 \\ 
RXTE/PCA    & 11.02$\pm$0.04 & 2.120$\pm$0.002 & 3.54          & 12.4 \\ 
NuSTAR      &  9.71$\pm$0.16 & 2.106$\pm$0.006 & 3.54          & 11.0 \\ 
HaloSat     & 10.20$\pm$0.17 & 2.12$\pm$0.03   & 3.54$\pm$0.13 & 11.5  \\ 
\tableline
\end{tabular}
\end{center}
\tablecomments{Normalization is in units of $\rm photons \, keV^{-1} \, cm^{-2} \, s^{-1}$ at 1~keV and $N_H$ is in units of $10^{21} \rm cm^{-2}$. Flux is in the 0.5-2 keV band in units of $(10^{-9} \rm \, erg \, cm^{-2} \, s^{-1})$. References are: \citet{Toor1974} for the historical instruments, \citet{Kirsch2005} for the PCA, and \citet{Madsen2017} for NuSTAR.}
\end{table}

We used the Crab to check the effective area of HaloSat.  The Crab is often used as a `standard candle' in X-ray astronomy \citep{Toor1974,Jahoda2006,Madsen2015}. However, it does exhibit variability of up to 7\% in the 10-100 keV band on long time scales \citep{Wilson2011}.

We extracted Crab spectra for each detector, see Fig.\ref{fig:crab_spec}, and applied the temperature-dependent energy calibration and response matrices discussed previously. We modeled the Crab spectrum as an absorbed powerlaw with the TBabs model in Xspec to describe the interstellar absorption \citep{Wilms2000}. The same absorption column density, $N_H$, and powerlaw photon index, $\Gamma$, were used for all detectors, but the normalization was allowed to vary between the detectors. We included a powerlaw with photon index and normalization fixed to the values found by \citet{Cappelluti2017} to model the cosmic X-ray background (CXB) subject to absorption with the TBabs model with the column density fixed to the average Galactic absorption within a $5\arcdeg$ radius of the line of sight \citep{HI4PI2016}. We added a powerlaw not modified by the response matrix to model the instrumental background. The instrumental background photon index was fixed to 0.65 and the normalization was allowed to vary between the detectors. We fitted the spectra in the 0.4-7~keV band.

We obtained a good fit with $\chi^2$/DoF = 1227.5/979 for $N_H = (3.54 \pm 0.13) \times 10^{21} \rm cm^{-2}$ and $\Gamma = 2.12 \pm 0.03$ (90\% confidence). The photon index is consistent with those measured for the Crab with {\it XMM-Newton}, the Proportional Counter Array (PCA) on the {\em Rossi X-ray Timing Explorer} \citep{Kirsch2005}, and NuSTAR \citep{Madsen2017}, see Table~\ref{tab:crab_flux}. We note that the photon index depends on the energy band used for the fitting. If we use a softer band, we find a harder photon index, consistent with the values reported for instruments sensitive in softer bands such as the Low Energy Concentrator Spectrometer on BeppoSAX and the Position Sensitive Proportional Counter on ROSAT \citep{Kirsch2005}. This suggests that a single power law does not provide an adequate representation of the spectrum of the Crab pulsar plus nebula at low energies, consistent with the observations that the pulsed flux is harder than the nebular emission.

The science of HaloSat is focused on emission in the 0.5-2~keV band. The Crab flux in that band depends on all of the model parameters, so we prefer to directly compare observed fluxes rather than only the powerlaw normalization. The HaloSat fluxes from the Crab in the 0.5-2~keV band in units of $10^{-9} \rm \, erg \, cm^{-2} \, s^{-1}$ are $11.54 \pm 0.20$ for DPU 14, $11.74 \pm 0.21$ for DPU 54, and $11.27 \pm 0.20$ for DPU 38 (uncertainties are 90\% confidence). The fluxes are consistent within the statistical error. The CXB contributes 4\% of the total counts in the 0.5-2~keV band while the instrumental background contributes 8\% to 10\% depending on the detector. Thus, uncertainty in modeling the instrumental background may contribute up to a few per cent uncertainty in measuring the Crab flux.

We compare the HaloSat Crab flux in the 0.5-2~keV band with the flux calculated from the spectral parameters, $\Gamma$ and normalization, measured with other instruments in Table~\ref{tab:crab_flux}. \citet{Madsen2017} recently measured the absolute flux of the Crab using a NuSTAR observation in which the detectors were illuminated directly without the X-rays passing through the optics. This greatly simplifies the instrument response. They found that the true Crab flux in the 3-7~keV band is $\sim$12\% higher than their previous choice for the Crab normalization based on results from contemporary missions with X-ray optics. This motivates our choice of collimated instruments for comparison. Because those measurements do not extend to the soft band needed to accurately measure the absorption, we use the absorption measured by HaloSat. We find that the Crab flux as measured by HaloSat agrees within 8\% with that inferred from the PCA spectral parameters and within 5\% with that inferred from the NuSTAR parameters and the average of historical spectral parameters in \citet{Toor1974}. Due to the uncertainty in extrapolating these measurements to the soft X-ray band and the simplicity of the response of HaloSat, we choose not to make any adjustments to the effective area of HaloSat.

Most of the previously published results on Milky Way halo emission use {\it XMM-Newton} which has two imaging instruments, the EPIC-MOS and the EPIC-pn. Unfortunately, the MOS suffers from pileup during observations of the Crab and the only Crab normalizations reported are for the EPIC-pn. Using the parameters of the Crab spectrum measured by \citet{Kirsch2005} for the EPIC-pn using Wilms abundances and Verner cross-sections, we find an absorbed flux of $(9.185 \pm 0.042) \times 10^{-9} \rm \, erg \, cm^{-2} \, s^{-1}$ in the 0.5-2~keV band. Cross calibrating the MOS and PN, \citet{Mateos2009} found that the MOS registers higher flux than the pn with the ratio being energy dependent. Averaging their values for the ratio of pn versus MOS1 and MOS2 in the 0.5-1 and 1-2~keV energy bands, we find a correction factor for the 0.5-2~keV band of 8.4\%. Comparing their NuSTAR absolute Crab flux measurement with the EPIC-MOS in the 3-7~keV band, \citet{Madsen2017} found that the MOS flux was 11.6\% lower. Combining these two factors, we conclude that the Crab pn flux should be corrected by a factor of 1.21, bringing the flux to $11.1 \times 10^{-9} \rm \, erg \, cm^{-2} \, s^{-1}$.  This is about 4\% lower than the HaloSat Crab flux. This is reasonable agreement given the uncertainties.


\subsection{Background}

\begin{figure}[tb]
\includegraphics[width=3.25in]{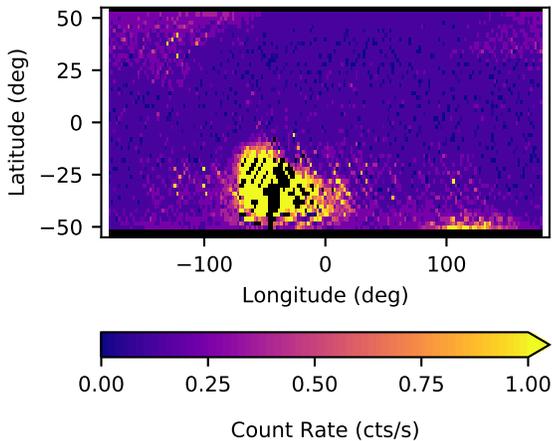}
\caption{Median count rate before background filtering versus orbital position for DPU 38 in the energy band 0.4-3~keV. The SAA is prominent and high background regions at high latitudes are visible.
\label{fig:orbit_background}}
\end{figure}

HaloSat was deployed from the ISS and has an orbital inclination of 51.6$\arcdeg$ which brings the spacecraft into regions of high background at the high and low latitude regions of the orbit as well as in the South Atlantic Anamoly (SAA), see Fig.~\ref{fig:orbit_background}. The SAA covers a well defined region and has a consistently high particle background. Instrument event processing is automatically turned off upon entrance to the SAA and resumed upon exit. The background at all latitudes is time variable. Data are collected in these regions and filters are applied in the data analysis to select times when the instrumental background is low. Optimization of the filtering algorithms and background modeling is currently underway.

We note that the orbital inclination of the ISS is not optimal for HaloSat because of the high instrumental background experienced mostly at high latitudes. Development of the capability for routine small satellite launches at lower inclinations of 30$\arcdeg$ or less would benefit future high energy astrophysics small satellite missions by providing lower instrumental background and increased observational efficiency.

\section{Initial Halo Observations} \label{sec:halo}

\begin{figure}[tb]
\includegraphics[width=3.25in]{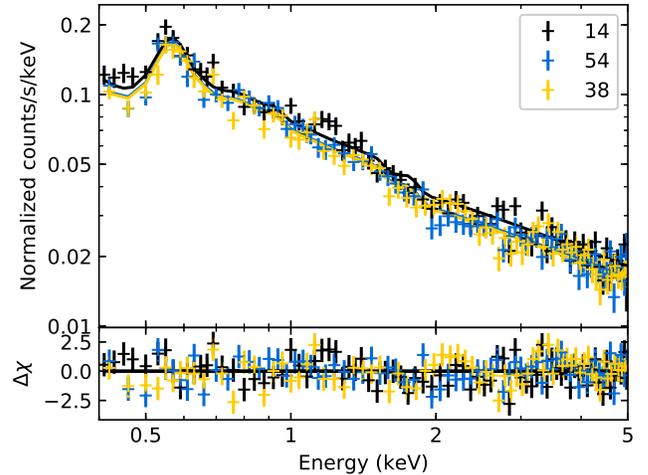}
\caption{X-ray spectra of a halo field. Data from all three detectors are shown as indicated by the legend.
\label{fig:halo_spec}}
\end{figure}

To make a useful contribution to the study of the hot halo of the Milky Way, HaloSat must achieve its design sensitivity for oxygen line emission and it must survey a large fraction of the high Galactic latitude sky. Figure~\ref{fig:halo_spec} shows a spectrum obtained for a high Galactic latitude field at $(l = 165.59\arcdeg, b = 61.92\arcdeg)$. We also analyzed data for a second field at $(229.25\arcdeg, 67.50\arcdeg)$ The temperature-dependent energy calibration described above was applied. 

Following \citet{Henley2012}, we fit the data with a model consisting of Gaussians with centroids fixed at 568.4~eV and 653.7~eV for the \ion{O}{7} and \ion{O}{8} line emission and an absorbed thermal plasma model (APEC) for emission from the halo and an unabsorbed APEC model for emission from the local bubble (LB). We fixed the temperature and emission measure for the local bubble APEC model using the results from \citet{Liu2017}. We removed the oxygen line emission from the APEC models following the procedure of \citet{Lei2009}, but only removing lines in the energy range from 0.484 to 0.744~eV which covers that modeled with the two Gaussians. We included a powerlaw with photon index fixed to 1.45 to model the CXB \citep{Cappelluti2017}. The halo APEC component and the CXB powerlaw component were subject to absorption with the TBabs model with the column density fixed to the average Galactic absorption within a $5\arcdeg$ radius of the line of sight \citep{HI4PI2016}. We added a powerlaw not modified by the response matrix to model the instrumental background.  The instrumental background photon index was fixed to 0.65 and the normalization was allowed to vary between the detectors. We used the C-statistic to perform fitting and the $\chi^2$-statistic to evaluate the quality of fit. Fitting over the 0.4-5~keV band.

\begin{table}[tb]
\begin{center}
\caption{Halo spectrum model parameters.\label{tab:halo_spec}}
\begin{tabular}{lcc} \tableline
Parameter                              & (229.25, 67.50)  & (165.59, 61.92)  \\ \tableline
Exposure (ks)                          & 42.6, 42.2, 41.9 & 44.4, 44.0, 42.7 \\
\ion{O}{7} flux (LU)                   & 5.14$\pm$0.39    & 4.33$\pm$0.35    \\ 
\ion{O}{8} flux (LU)                   & 0.94$\pm$0.20    & 0.90$\pm$0.18    \\ 
$N_{H}$  $(\rm 10^{20} \, cm^{-2})$    & 1.41             & 1.13             \\
Halo $kT$ (keV)                        & 0.178$\pm$0.007  & 0.189$\pm$0.010  \\
Halo EM ($10^{-3}$ \, cm$^{-6}$ \, pc) & 25.8$\pm$4.3     & 19.8$\pm$3.8     \\
LB $kT$ (keV)                          & 0.097            & 0.097            \\
LB EM                                  & 4.14             & 4.70             \\
CXB $\Gamma$                           & 1.45             & 1.45             \\
CXB norm                               & 10.3$\pm$0.7     & 9.3$\pm$0.6      \\
$\chi^2$/DoF                           & 776.1/679        & 799.64/679       \\ \tableline 
\tableline
\end{tabular}
\end{center}
\tablecomments{Uncertainties are 1$\sigma$ confidence. Quantities without uncertainties were held fixed. The exposures are after background and data quality screening for DPU 14, 54, and 38. The CXB normalization is at 1~keV with units of $\rm keV \, cm^{-2} \, s^{-1} \, sr^{-1} \, keV^{-1}$.}
\end{table}

The fit parameters are shown in Table~\ref{tab:halo_spec}. The statistical accuracy on the \ion{O}{7} flux meets our observational goal. The \ion{O}{7} flux for the field at $(l = 165.59\arcdeg, b = 61.92\arcdeg)$ is lower than the fluxes measured by \citet{Henley2012} using {\it XMM-Newton} for lines of sight within the HaloSat field which range from 5.34$\pm$0.44~LU to to 8.47$\pm$0.24~LU.  The flux for the field at $(l=  230.1\arcdeg, b= 66.2\arcdeg)$ is slightly higher than the fluxes of 3.54$\pm$0.36~LU and 3.16$\pm$0.28~LU from \citet{Henley2012} in the same field. The CXB normalizations are in reasonable agreement with that of \citet{Cappelluti2017}. 


\begin{figure*}[tb]
\centerline{\includegraphics[width=6.75in]{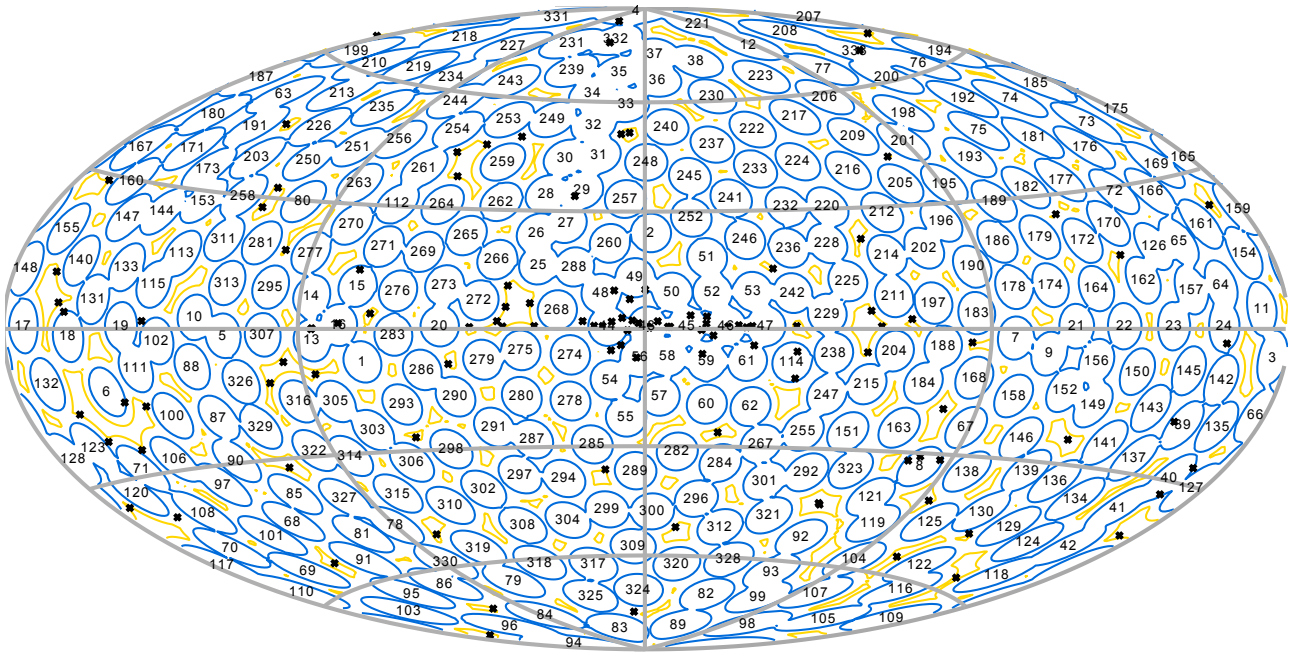}}
\caption{HaloSat targets in Galactic coordinates. The Galactic center is at the center of the image, longitude increases to the left, and latitude increases toward the top. The blue contours show the 10$\arcdeg$diameter full-response field of view for each target and the yellow contours show the 14$\arcdeg$ diameter zero response. The black X’s show bright X-ray sources.
\label{fig:targets}}
\end{figure*}

\begin{deluxetable}{rlccccc}
\tabletypesize{\footnotesize}
\tablecaption{HaloSat targets \label{tab:targets}}
\tablehead{\colhead{ID} & \colhead{Name} & \colhead{Type} & \colhead{RA} & \colhead{DEC} & \colhead{$l$} & \colhead{$b$}}
\startdata
  1 & Cygnus Loop       & C & 312.75 &  30.67 &  73.98 &  -8.56 \\
  2 & Sco X-1           & C & 244.98 & -15.64 & 359.09 &  23.78 \\
  3 & Crab              & C &  83.63 &  22.01 & 184.56 &  -5.78 \\
    & \ldots \\
  8 & LMC               & S &  78.83 & -67.78 & 278.32 & -34.00 \\
  9 & Pup A offset      & S & 118.07 & -39.33 & 254.27 &  -6.22 \\
 10 & Tycho SNR         & S &   5.79 &  64.66 & 119.91 &   1.95 \\
    & \ldots \\
 66 & HSWCX1            & X &  73.27 &  17.51 & 182.73 & -16.37 \\
 67 & HSWCX2            & X & 101.92 & -61.29 & 271.21 & -24.01 \\
 68 & MSWCX1            & X &  15.46 &  18.24 & 126.40 & -44.56 \\
    & \ldots \\
 80 & HALO J1807+699    & H & 271.83 &  69.98 & 100.31 &  29.12 \\
 81 & HALO J0011+119    & H &   2.86 &  11.94 & 107.69 & -49.75 \\
 82 & HALO J0021-440    & H &   5.45 & -44.07 & 320.42 & -72.04 \\
 83 & HALO J0021-255    & H &   5.44 & -25.57 &  44.48 & -83.17 \\
    & \ldots \\
\enddata
\tablecomments{Type: C = Calibration, S = Secondary science, X = Solar wind charge exchange, H = Halo. RA and DEC are J2000 coordinates. $l$ and $b$ are Galactic coordinates.  This table is published in its entirety in the machine-readable format. A portion is shown here for guidance regarding its form and content.}
\end{deluxetable}

In order to measure the properties of the halo, we must conduct similar observations over a large fraction of the sky. Our goal is to survey the entire sky, although the fields with $|b| > 30\arcdeg$ will be most useful in constraining the properties of the halo. The size of the FoV determines the number of targets needed to cover the sky. We selected 333 targets to tile the sky including targets selected for instrument calibration, SWCX studies, and secondary science on bright, extended soft X-ray sources. The positions of the halo targets were chosen to minimize overlap and avoid bright X-ray sources in the ROSAT, Uhuru, and MAXI catalogs, particularly those with low-energy line features. The targets are presented in Table~\ref{tab:targets} and shown in Galactic coordinates in Fig~\ref{fig:targets}.

\section{Conclusions}

The initial results presented here demonstrate that HaloSat should help advance our understanding of the hot halo of the Milky Way and provide a unique data set for the study of solar wind charge exchange emission. Thus, CubeSats can be effective vehicles for astrophysics research even within their limited mass, power, and volume constraints and be constructed and operated at modest cost by exploiting the commercialization of small satellite technologies. The success of HaloSat should encourage construction of more CubeSats for astrophysics.

\acknowledgments

We acknowledge support from NASA grant NNX15AU57G. We thank Steve Schneider for leading the bus work, Tracy Behrens for her tireless work in assembling the HaloSat electronics, Luis Santos for system engineering, Keith White for mechanical design and drawing, Chris Esser for spacecraft mechanical design, Tom Golden and Doug Laczkowksi for mission operations, Rich Dvorsky for mechanical and polymerics advice, and Jeff Dolan for advice about connectors and electronics. 

We are grateful to Brian Busch, Larry Detweiler, and Matt Miller for machining parts, Kayla Racinowski for help with polymerics, Jim Phillips for help with inductor cores, Jesse Haworth for HVPS testing and designing the logo, Riley Wearmouth for early mechanical designs, Mike Matthews for work on the concept of operations and communications, Brenda Dingwall for program management, Calvin Whitaker and Tyler Roth for system administration, Tim Cameron for building the HVPS, and John Hudeck for seeing us through instrument vibration testing and for mentoring Keith White.

Our reviewers, Dave Sheppard, Scott Porter, Kevin Black, and Jasper Halekas, provided an invaluable service. Kristen Hanslik, Charles Dumont, Karl Hansen, Bruce Patterson, Natali Vannoy, Jake Beckner, David Hall, Jesse Ellison, and Matt Pallas did a great job seeing us through integration and testing, as did Scott Inlow on thermal modeling, Matt Baumgart and Bryan Rogler on GNC engineering, Nick Monahan, Austin Bullard, and Jeff Adams on mission operations, Rebecca Walter on RF engineering, and Steve Bundick and Matt Schneider on communications testing. 

We couldn't talk to HaloSat without the WFF UHF Ground Station Team and their continuing support. We thank Devon Sanders, Ned Riedel, Allen Crane, Larry Madison, and Kyle McLean for their work on alignment, Dan Evans and Mike Garcia for help from above, and Tristan Prejean, Conor Brown, and many others at Nanoracks for getting HaloSat into orbit. We probably wouldn't have been selected without the work of Ben Cervantes, Will Mast, Scott Schaire, Ryoichi Hasebe, Brad Hood, Sally Smith, and Brooks Flaherty on the Mission Planning Laboratory run for the proposal.

%

\vspace{5mm}

\facility{HaloSat}
\software{XSPEC \citep{Arnaud1996}, astropy \citep{astropy}}




\begin{thebibliography}{}

\bibitem[Aschenbach, B., Egger, R., \& Tr\"umper()]{Aschenbach1995} Aschenbach, B., Egger, R., \& Tr\"umper, J., “Discovery of Explosion Fragments Outside the Vela Supernova Remnant Shock-Wave Boundary”, Nature, 373, 587, 1995.


\bibitem[Arnaud(1996)]{Arnaud1996} Arnaud, K.A. 1996, ASP Conf., 101, 17 

\bibitem[Astropy Collaboration et al.(2013)]{astropy} Astropy Collaboration, Robitaille, T.~P., Tollerud, E.~J., et al.\ 2013, \aap, 558, A33 

\bibitem[Becker et al.(1979)]{Becker1979} Becker, R.H., Holt, S.S., Smith, B.W. et al. 1979, ApJ, 234, L73 

\bibitem[Boylan-Kolchin et al.(2013)]{Boylan2013} Boylan-Kolchin, M., Bullock, J. S., Sohn, S. T. et al. 2013, ApJ, 768, 140 

\bibitem[Cappelluti et al.(2017)]{Cappelluti2017} Cappelluti, N., Li, Y., Ricarte, A., et al. 2017, ApJ, 837, 19 

\bibitem[Cen \& Ostriker(1999)]{Cen1999} Cen, R. and Ostriker, J.P. 1999, ApJ, 514, 1 

\bibitem[Crusan \& Galcia(2019)]{Crusan2019} Crusan, J.\ \& Galica, C.\ 2019, Acta Astronautica, 157, 51 

\bibitem[Dame(1993)]{Dame1993} Dame, T.M. 1993, AIP Conf. Ser., 278, 267

\bibitem[Galeazzi et al.(2014)]{Galeazzi2014} Galeazzi, M., Chiao, M., Collier, M.R. et al. 2014, Nature, 512, 171 

\bibitem[Gendreau et al.(2016)]{Gendreau2016} Gendreau, K.C., Arzoumanian, Z., Adkins, P. W. et al. 2016, Proc. SPIE, 9905, 99051H 

\bibitem[Grcevich \& Putnam(2009)]{Grcevich2009} Grcevich, J. \& Putnam, M. E. 2009, ApJ, 696, 385 

\bibitem[Gupta et al.(2012)]{Gupta2012} Gupta, A., Mathur, S., Krongold, Y., Nicastro, F., and Galeazzi, M., ApJ, 756, L8 

\bibitem[Hegel(2016)]{Hegel2016} Hegel, D. 2016, in 30th Annual AIAA/USU Conference on Small Satellites, SSC16-X-7, \url{https://digitalcommons.usu.edu/smallsat/2016/TS10AdvTech2/5/} 

\bibitem[Henley et al.(2010)]{Henley2010} Henley, D.B., Shelton, R.L., Kwak, K., Joung, M. R., Mac Low, M.-M. 2010, ApJ, 723, 935 

\bibitem[Henley \& Shelton(2012)]{Henley2012} Henley, D.B. \& Shelton, R.L. 2012, ApJS, 202, 14 

\bibitem[Hevner et al.(2011)]{Hevner2011} Hevner, R., Holemans, W., Puig-Suari, J., Twiggs, R.\ 2011, Proceedings of the AIAA/USU Conference on Small Satellites, From 0 to 7.5 km/s, SSC11-II-3, \url{https://digitalcommons.usu.edu/smallsat/2011/all2011/15/}

\bibitem[HI4PI Collaboration et al.(2016)]{HI4PI2016} HI4PI Collaboration, Ben Bekhti, N., Fl\"oer, L., et al.\ 2016, \aap, 594, A116

\bibitem[Holt et al.(1994)]{Holt1994} Holt, S.S., Gotthelf, E.V., Tsunemi, H., Negoro, H. 1994, PASJ, 46, L151 

\bibitem[Jahoda et al.(2006)]{Jahoda2006} Jahoda, K., Markwardt, C.B., Radeva, Y. et al. 2006, ApJS, 163, 401 

\bibitem[Kaaret et al.(2019)]{Kaaret2019} Kaaret, P., Zajczyk, A., LaRocca, D. et al. 2019, Proceedings of the AIAA/USU Conference on Small Satellites, Year in Review I, SSC19-III-05, \url{https://digitalcommons.usu.edu/smallsat/2019/all2019/277/}

\bibitem[Kirsch et al.(2005)]{Kirsch2005} Kirsch, M.G., Briel, U.G., Burrows, D. et al. 2005, Proc. SPIE, 5898, 22 

\bibitem[Koutroumpa et al.(2009)]{Koutroumpa2009} Koutroumpa, D., Collier, M.R., Kuntz, K.D. et al. 2009, ApJ, 697, 1214 

\bibitem[Kuntz(2019)]{Kuntz2019} Kuntz, K.D. 2019, A\&AR, 27, 1 

\bibitem[Lei et al.(2009)]{Lei2009} Lei, S., Shelton, R. L., Henley, D. B. 2009, ApJ, 699, 1891 

\bibitem[Liu et al.(2017)]{Liu2017} Liu, W., Chiao, M., Collier, M. R. et al. 2017, ApJ, 834, 33 

\bibitem[Madsen et al.(2015)]{Madsen2015} Madsen, K.K., Harrison, F.A., Markwardt, C.B. et al. 2015, ApJS, 220, 8 

\bibitem[Madsen et al.(2017)]{Madsen2017} Madsen, K.K., Forster, K., Grefenstette, B.W., Harrion, F.A., Stern, D. 2017, ApJ, 841, 56 

\bibitem[Mateos et al.(2009)]{Mateos2009} Mateos, S., Saxton, R.D., read, A.M., Sumbay, S. 2009, A\&A, 496, 879 

\bibitem[McMillan(2011)]{McMillan2011} McMillan, P.J. 2011, MNRAS, 414, 2446-2457 

\bibitem[Miller \& Bregman(2015)]{Miller2015} Miller, M. J. \& Bregman, J. N. 2015, ApJ, 800, 14 

\bibitem[Moretti et al.(2009)]{Moretti2009} Moretti, A., Pagani, C., Cusumano, G. et al. 2009, A\&A, 493, 501 

\bibitem[Nicastro et al.(2002)]{Nicastro2002} Nicastro, F., Zezas, A., Drake, J.\ et al. 2002, ApJ, 573, 157 

\bibitem[Planck Collaboration(2014)]{Planck2014} Planck Collaboration, 2014, \aap,  571, A16 

\bibitem[Revnivtsev et al.(2005)]{Revnivtsev2005} Revnivtsev, M., Gilfanov, M., Jahoda, K., Sunyaev, R. 2005, \aap, 444, 381 

\bibitem[Scholze \& Procop(2009)]{Scholze2009} Scholze, F. \& Procop, M. 2009, X-Ray Spectrometry, 38, 312321 

\bibitem[Shull, Smith,& Danforth(2012)]{Shull2012} Shull, J.M., Smith, B.D., \& Danforth, C.W. 2012, ApJ, 759, 23 

\bibitem[Slavin et al.(2013)]{Slavin2013} Slavin, J., Wargelin, B.J., \& Koutroumpa, D. 2013, ApJ, 779, 13 

\bibitem[Smith et al.(2007)]{Smith2007} Smith, R.K., Bautz, M. W., Edgar et al. 2007, PASJ, 59, 141 

\bibitem[Snowden et al.(1997)]{Snowden1997} Snowden, S.L., Egger, R., Freyberg, M. J. et al. 1997, ApJ, 485, 125 

\bibitem[Spitzer(1956)]{Spitzer1956} Spitzer, L. 1956, ApJ, 124, 20 

\bibitem[Toor \& Seward(1974)]{Toor1974} Toor, A. \& Seward, F. D. 1974, AJ, 79, 995 

\bibitem[Weisskopf et al.(2010)]{Weisskopf2010} Weisskopf, M.C., Guainazzi, M., Jahoda, K. et al. 2010, ApJ, 713, 912 

\bibitem[Wilms, Allen, \& McCray(2000)]{Wilms2000} Wilms, J., Allen, A., \& McCray, R. 2000, ApJ, 542, 914 

\bibitem[Wilson-Hodge et al.(2011)]{Wilson2011} Wilson-Hodge, C.A., Cherry, M.L., Case, G.L. et al. 2011, ApJ, 727, L40 

\bibitem[Yoshino et al.(2009)]{Yoshino2009} Yoshino, T., Mitsuda, K., Yamasaki, N. Y. et al. 2009, PASJ, 61, 805 

\bibitem[Zajczyk et al.(2018)]{Zajczyk2018} Zajczyk, A., Kaaret, P., Kirchner, D.L. et al. 2018, Proceedings of the AIAA/USU Conference on Small Satellites, Upcoming Missions, SSC18-WKIX-01, \url{https://digitalcommons.usu.edu/smallsat/2018/all2018/471/}

\bibitem[Zajczyk et al.(2019)]{Zajczyk2019} Zajczyk, A., Kaaret, P. et al. 2019, in preparation.

\bibitem[Zhezher et al.(2017)]{Zhezher2017} Zhezher, Ya. V., Nugaev, E. Ya., \& Rubtsov, G. I. 2017, Astron. L., 42, 173-181 


\end{thebibliography}
\end{document}